\documentclass[prl,twocolumn]{revtex4}
\usepackage{graphicx,epsfig}
\newcommand{\eps}{\varepsilon}

\newcommand{\A}{\mathcal{A}}
\newcommand{\Tr}{\mathrm{Tr}}

\newcommand{\dbar}{\kern-.1em{\raise.8ex\hbox{ -}}\kern-.6em{d}}

\begin{document}
\input epsf
\title{Optimal rotations of deformable bodies \\ and orbits in magnetic fields}%
\author{ J.E. Avron,
O. Gat, O. Kenneth and U. Sivan}
\address{Department of physics, Technion, Haifa 32000, Israel
}%
\email{avron@physics.technion.ac.il}

\date{\today}%
% ----------------------------------------------------------------

% ----------------------------------------------------------------

\begin{abstract}
Deformations can induce rotation with zero angular momentum where
dissipation is a natural ``cost function''. This gives rise to an
optimization problem of finding the most effective rotation with
zero angular momentum. For certain plastic and viscous media in
two dimensions the optimal path is the orbit of a charged particle
on a surface of constant negative curvature with magnetic field
whose total flux is half a quantum unit.
\end{abstract}
\maketitle

Rotations with zero angular momentum are intriguing. The most
celebrated phenomenon of this kind is the rotation of a falling
cat. A mechanical model \cite{ref:kane} replacing the cat by two
rigid bodies that can rotate relative to each other, has been
extensively studied, see \cite{ref:montgomery,ref:marsden} and
references therein. Here we address rotations with zero angular
momentum under linear deformations. Our motivation comes from
nano-mechanics: Imagine an elastic or plastic material with its
own energy source, and ask what is the most efficient way of
turning it through an appropriate sequence of autonomous
deformations without external torque.

Deformations can generate rotations because order matters: A cycle
of deformations will, in general, result in a rotation. The
limiting ratio between the rotation and the area of the controls
(deformations), when the latter tends to zero can be interpreted
as curvature \cite{ref:novikov,ref:shapere}. Consequently, small
cycles are ineffective since a cycle of length $\eps$ in the
controls leads to a rotation of order $\eps^2$. The search for
optimal paths forces one to mind deformations that are not small.

The problem we address has three parts. The first part is to
determine the rotation for a given path of the controls. We solve
this problem for general linear deformations. In two dimensions
this leads to curvature on the space of the controls which is
localized. If one thinks of the curvature as a magnetic field,
then the total magnetic flux is  half the quantum unit. The second
part is to set up a model for the cost function which we choose to
be a measure of dissipation. We focus on two settings, one where
the dissipation is rate independent, as is the case in certain
plastics, and the other where it is rate dependent as in liquids.
Both cost functions lead to the same metric on the space of
deformations. The third part is to pose and solve the problem of
finding the path of minimal dissipation for a given rotation. In
two dimensions and for either model of dissipation, the problem
maps to finding the shortest path that starts at a given point and
encloses a given amount magnetic flux. As we shall see, optimal
paths tend to linger near the circle in the space of controls
where the ratio of eigenvalues of the quadrupole moment of the
body is $(\phi+\sqrt{\phi})^2$. $\phi=(1+\sqrt{5})/2$ is the
golden ratio.

Deformations generate rotations {\em because} angular momentum is
conserved \cite{ref:shapere}. Consider a collection of point
masses $m_\alpha$ at positions $x_\alpha$. Internal forces may
deform the body, but there are no external forces. Suppose that
the center of mass of the body is at rest at the origin and that
the body has zero angular momentum. The total angular momentum,
$L_{ij}$, must then stay zero for all times.

A linear deformation is represented by a matrix $M$ that sends
$x\to M x$.  The $i,j$ component of the angular momentum is
\begin{equation}L_{ij}
=\Tr\,(\dot M Q M^t\, \ell_{ij}),%\nonumber
\end{equation}
where $Q$ is the quadrupole moment of the body
\begin{equation}\label{eq:qmoment}
Q_{ij}=\sum_\alpha m_\alpha (x_\alpha)_i(x_\alpha)_j
\end{equation}  and  $\ell_{ij}, \ i<j$ are the $n\choose 2$
generators of rotations in $n$ dimensions, {\it i.e.}
$(x,\ell_{ij}y)=x_iy_j-x_jy_i$. Since $\ell_{ij}$ span the
anti-symmetric matrices, the set of  $n\choose 2$ equations
$L_{ij}=0$  imply that the matrix $(dM)QM^t$ is symmetric.
%%%%%%%%%%%%%%%%%%%%

Two immediate consequences of this symmetry are:
\begin{itemize}
\item Isotropic bodies:  With $Q=1$, and $M$ close to the identity,
the symmetry of $(dM)QM^t$ reduces to $dM$ being symmetric: the
linear transformation must be a strain \cite{ref:landau}.
%%%%%%%%%%%%%%%%%%%%%
\item Pointers: Pointers are  bodies with large aspect ratios,
such as needles and discs. In the limit of infinite aspect ratio,
$Q$ may be identified with a projection where $\dim Q$ is the
dimension of the pointer. With $M$ near the identity, the symmetry
of $ (dM) Q$ implies that $ (1-Q)(dM) Q=(1-Q) Q (dM)^t = 0. $
Since $Q$ does not acquire a component in the normal direction,
$1-Q$, under $dM$ a pointer keeps its orientation.
\end{itemize}
%%%%%%%%%%%%%%%%%%%%%%%%

We now derive the fundamental relation between the response
(rotation) and the controls (deformations). To this end we use the
polar decomposition $M=RS$ with $R$ orthogonal and $S$ positive.
Assuming $S$ positive is a choice of a gauge which makes the
representation unique with $S=\sqrt{M^tM}$. The symmetry of
$(dM)QM^t$ gives
 \begin{equation}\label{eq:CAR0}\{
\A,SQS\} =SQ(dS)-(dS)QS, \quad \A=R^{-1}\dbar R.\end{equation}
Eq.~(\ref{eq:CAR0})  determines the  differential rotation, $\A$,
in terms of the variation of the controls $dS$.  The symbol
$\dbar$  stresses that the differential will not, in general,
integrate to a function on the space of deformations.
Geometrically, the differential rotation is the connection 1-form
which fixes a notion of parallel transport.

Eq.~(\ref{eq:CAR0}) can be interpreted in terms of a variational
principle: The motion is such that the kinetic energy is minimal
for a given deformation. To see this, let $M=1$ and
$\dot M =\dot R +\dot S $
with $\dot R$ antisymmetric ({\it i.e.} a rotation) and $\dot S$
symmetric ({\it i.e.} a strain). The kinetic energy is
\begin{equation}\label{eq:kinetic}
E=\frac 1 2 \Tr\,\big( \dot M Q\dot M^t\big) =\frac 1
2\,\Tr\,\big( (\dot R+\dot S) Q (-\dot R+\dot S)\big)\; .
\end{equation}
Minimizing with respect to $\dot R$ gives
\begin{equation}
0=\delta E= \frac 1 2 \Tr\,\big( \delta \dot R \big(-\{Q,\dot
R\}+[ Q,\dot S]\big)\; .
\end{equation}
The trace is of a product of antisymmetric matrices, and its
vanishing for an arbitrary antisymmetric $\delta \dot R$ implies
$\{\dot R,Q\}=[ Q,\dot S]$ which is Eq.~(\ref{eq:CAR0}) for $M=1$.

One readily sees that if $\A$ is the solution of Eq.~(\ref{eq:CAR0})
 given $S$ and $Q$, then it is also a solution for $\lambda
S$ and $Q$ for $\lambda$ a scalar valued function. Hence scaling
does not drive rotations and we may restrict ourselves to volume
(or area) preserving deformations with $\det S=1$.

Since any $Q$ is obtainable by a linear deformation of the
identity, we may assume without loss of generality that $Q=1$.
Eq.~(\ref{eq:CAR0}) reduces to
 \begin{equation}\label{eq:CAR}
\{\A,S^2\}=[S,dS]
 \; .
\end{equation}
 Eq.~(\ref{eq:CAR})  is
conveniently solved in a basis where $S$ is diagonal. Let $s_j$
denote the eigenvalues of $S$ then
\begin{equation}\label{eq:connection}
\A_{ij}=\frac {s_i-s_j} {s_i^2+s_j^2}\,(dS)_{ij} \; .
\end{equation}
The curvature $F$ is more interesting than the connection. It is
defined by $F=d\A+\A\wedge \A$. Calculation gives
$$
F_{ij}=2\sum_k \frac{s_i\,s_j\, (s_i+s_k)(s_j+s_k)}{(s_i^2+s_j^2)
(s_j^2+s_k^2)(s_i^2+s_k^2)}\, (dS)_{ik}\wedge (dS)_{kj} \; .
$$

%%%%%%%%%%%%%%%%%%%%%%%%%%%%%%%%%%%%%%%%%%%%%%%%%%%%%%%%%%%%%%%%%%%%

The situation is particularly simple in two dimensions. We use the
Pauli matrices $\sigma_x, \sigma_z, i\sigma_y$ and the identity as
basis for the real $2\times 2$ matrices. The symmetric matrices
make a three dimensional space that is conveniently parameterized
by cylindrical coordinates:
\begin{equation}\label{eq:2-dS}
S(t,\rho,\theta )=t+ (\rho\cos\theta)\,\sigma_x+
(\rho\sin\theta)\,\sigma_z\,\quad  \rho\ge 0.
\end{equation}
In this parametrization $\det S=t^2-\rho^2$ and positive matrices
correspond to the cone $t>\rho$. Since $i\sigma_y$ is the
generator of rotation in two dimensions ${\cal
A}=-i\,\sigma_y\,\dbar\varphi$. Eq.~(\ref{eq:CAR}) can be readily
solved for the connection $\dbar\varphi$ whose differential is the
curvature $F$. One finds
\begin{equation}\label{eq:CAR2}
\dbar \varphi=\frac 1 {1+\zeta^2}\, d\theta,\quad F= -\frac 1
{(1+\zeta^2)^2}\,d(\zeta^2)\wedge d\theta \,,
\end{equation}
where $\zeta= t/\rho$.  $F$ and $\dbar\varphi$ are invariant under
scaling of $S$, as they must be.

The total curvature associated with the area preserving
deformations is then $\int_{t^2-\rho^2=1} F =\frac 1 2 \int
d\theta=\pi$. This implies that in any single closed cycle ({\it
i.e.} one without self intersections), the angle of rotation is at
most $\pi$. When interpreted as magnetic flux, $\pi$ corresponds
to half a unit of quantum flux.

A geometric understanding of why the total curvature of
orientation preserving deformations ($t>\rho$ in
Eq.~(\ref{eq:2-dS})) is $\pi$ comes from considering also the
orientation reversing deformations ($t<\rho$ in
Eq.~(\ref{eq:2-dS})). When both are considered $\zeta$ takes
values in $[0,\infty)$ and then $F$ coincides with the curvature
of the canonical line bundle (Berry's spin half) on the
(stereographically projected) two sphere. The Chern number of the
bundle is $1$ and the total curvature $2\pi$. The orientation
preserving matrices correspond to the northern hemisphere,
$\zeta>1$, and orientation reversing to the southern hemisphere,
$\zeta<1$, \cite{rem:plane}.

%%%%%%%%%%%%%%%%%%%%

The cost function must include some measure of dissipation. For,
without dissipation energy is a function of the controls and no
change in energy is associated with a closed loop. We consider two
models of dissipation in isotropic media. Both lead to the same
metric on the space of controls, namely
\begin{equation}\label{eq:inv-metric} (d\ell)^2=
\,\Tr\left(S^{-2}\,d(S^2)\otimes S^{-2}d(S^2)\right)\ .
\end{equation}

Consider a medium with viscosity tensor $\eta$. The power due to
dissipation is
\begin{equation}\label{eq:viscosity-dissipation}
P=  \frac 1 2\,\sum \eta^{ijkl}\dot u_{ij}\dot u_{kl}\; ,
\end{equation}
where $u$ is the strain tensor and $\dot u$ the strain rate. $S$
is related to $u$ by  $2u=S^2-1$. To see this recall that the
strain is {defined} as the change in the distance between two
neighboring points caused by a deformation. If the deformation is
described by a metric $g$ then $g=1+2u$, where $1$ is the metric
associated with the undeformed reference system~\cite{ref:landau}.
When considering {\em linear} deformation described by a symmetric
matrix $S$, the resulting metric is $g=S^2$ (regarding the
covariant components of $g$ as the elements of a positive matrix).

The space of (symmetric) 4-th rank isotropic tensors
is two dimensional and spanned by the two viscosity coefficients
$\eta$ and $\eta'$
 \cite{ref:landau}
\begin{equation}\label{eq:viscosity-scalar}
\eta^{ijkl}=\eta g^{ik}g^{jl}+\eta' g^{ij}g^{kl}\ .
\end{equation}
We therefore find that in an isotropic medium
\begin{equation}\label{eq:dissipation}
P= 2\eta\, Tr \left(g^{-1} \dot g g^{-1}\dot g\right)   +2\eta'\,
Tr\,\left( g^{-1}\dot g\right)Tr\,\left( g^{-1}\dot g\right)
\end{equation}
%%%%%
For volume preserving transformations the term multiplying $\eta'$
vanishes and one is left with the first term alone. By choosing
the unit of time appropriately one can take  $\eta=1$. This leads
to the metric of Eq.~(\ref{eq:inv-metric}) which is invariant
under congruence, $g\to AgA^t$, for an arbitrary invertible matrix
$A$.

The dissipation in certain plastic materials can be rate
independent. This is the continuum mechanics analog of the
dissipation due to friction when one body slides on another
\cite{ref:ori}. In plastics, rate independence  is a consequence
of the L\'evy-Mise constitutive relation: $s\,
\delta\lambda=\delta u$, where $s$ is the stress and $\delta
\lambda$ a scalar valued function \cite{ref:hill}. The
constitutive relation is formally the same as for fluids
\cite{ref:hill}, and by isotropy, the dissipation must be a
function of $d\ell$ of Eq.~(\ref{eq:inv-metric}). If the material
is memoryless, dissipation is additive with respect to
concatenating paths and must be proportional to $d\ell$.

%%%%%%%%%%%%%%%%%%%%%%%%%%%%%%%%%%%%%%%%%%%%%%%%%%
%%%%%%%%%%%%%%%%%%%%%%%
%%%%%%%%%%%%%%%%%%%%%%%%%%%%%%%%
Returning now to two dimensions, let us parameterize the area
preserving transformation by: $(t,\rho,\theta)=(\cosh(\tau/2),
\sinh(\tau/2),\theta)$. The metric Eq.~(\ref{eq:inv-metric}) gives
\begin{equation}\label{eq:metric-2d}
(d\ell)^2=(d\tau)^2+\sinh^2\tau(d\theta)^2\,.
\end{equation}
This metric gives the hyperboloid the geometry of the
pseudo-sphere i.e. it makes it into a surface of constant negative
curvature $-1$ \cite{ref:novikov}. Geometrically, this corresponds
to embedding the hyperboloid $t^2-\rho^2=1$ in Minkowski space.

\begin{figure}[h]
\includegraphics[width=4cm]{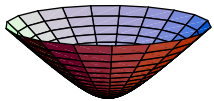}
\vskip-1.cm \caption{Positive, area preserving linear
transformation may be identified with the hyperboloid of
revolution $t^2-\rho^2=1$. Embedding the hyperboloid in Minkowski
space gives it the structure of the Lobachevsky plane with
constant negative curvature.}\label{fig:hyper}
\end{figure}

\begin{figure}[h]
\hbox{
\includegraphics[width=4.2cm]{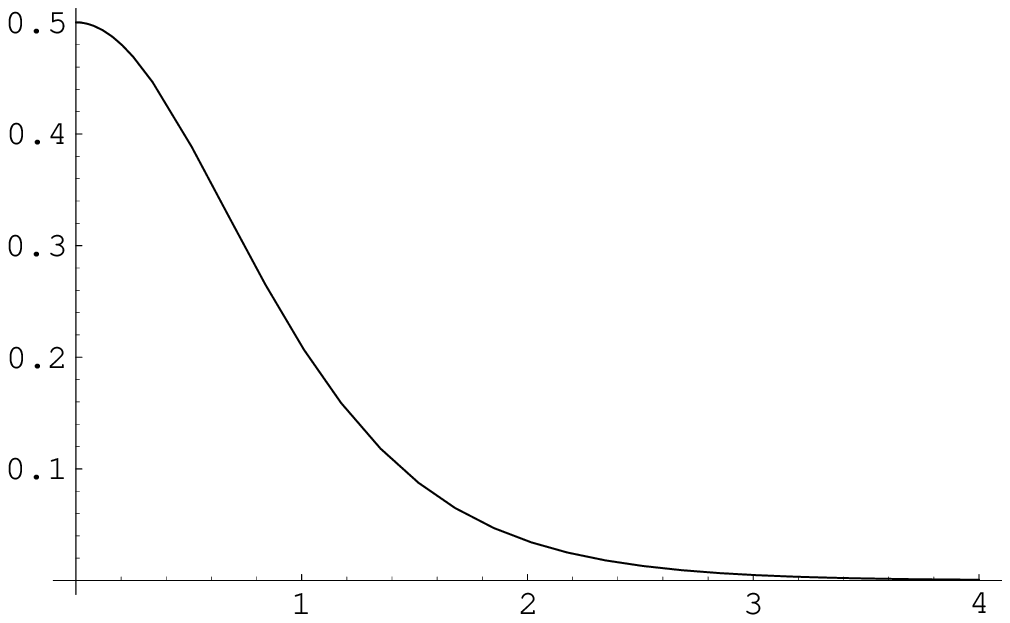}
\includegraphics[width=4.2cm]{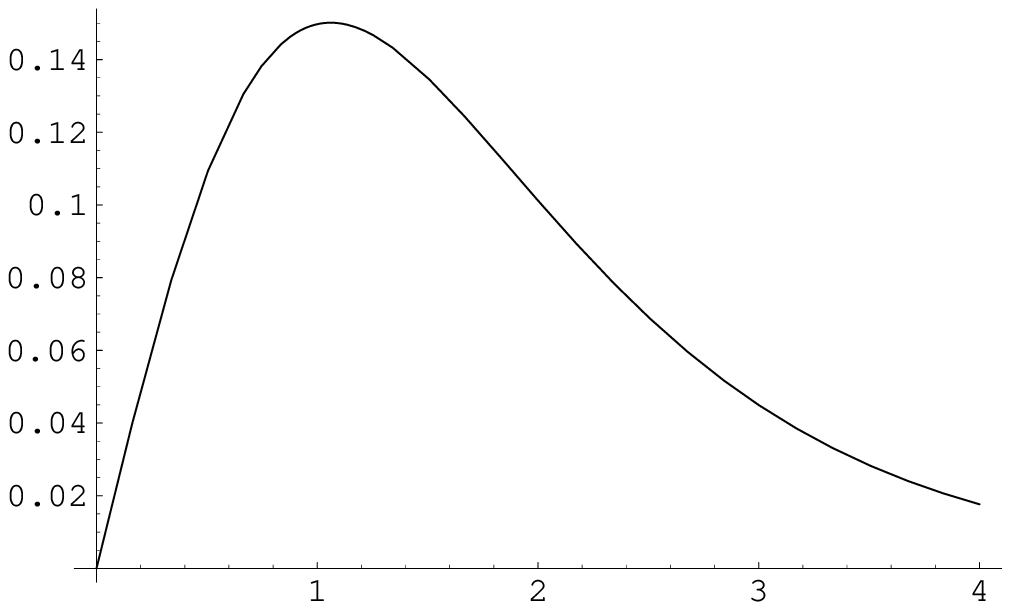}
} \caption{The scalar curvature, $f(\tau)$, which can be
interpreted as a magnetic field (left) and the $\theta$ component
of the vector potential $a(\tau)$. $f$ is exponentially localized
near the origin of the control space $(\tau,\theta)$, while $a$
has a maximum on the circle $\cosh\tau=\phi$, $\phi$ the golden
ratio. $a^2/2$ appears also as a potential in the effective
1-dimensional dynamics of the optimization problem.}
\label{fig:curv}
\end{figure}

The metric enables us to assign a scalar $f$ to the curvature
2-form $F$ of Eq.~(\ref{eq:CAR2}) as the ratio between $F$ and the
area form of the pseudo-sphere, $ \sinh\tau\, d\tau\wedge
d\theta$. Similarly, $a(\tau)$ is the scalar relating the length
form $\sinh\tau\, d\theta$ to the one-form $\dbar\varphi$. A
calculation then gives
\begin{equation}\label{eq:def-f-a} f(\tau)=\frac 1 {2\cosh^2\tau},
\quad a(\tau)=\frac{\cosh \tau -1}{\sinh 2\tau}
\end{equation} and is plotted in Fig. \ref{fig:curv}.
The curvature is everywhere positive and it is concentrated  near
the origin, $\tau=0$. It decays exponentially with $\tau$. This
means that large deformations are ineffective. We already know
that small deformations are ineffective. This brings us to the
optimization problem.

%%%%%%%%%%%%%%%%%%%%%%%%%%%%%%%%%%%%%%%%%%%%%%%%%%%
%%%%%%%%%%%%%%%%%%%%%%%%%%%%%%%%%%%%%%%%

The control problem is to find a closed path $\gamma$ in the space
of deformations, starting at $S_0=\sqrt Q_0$, ($Q_0$ is the
initial quadrupole),  which rotates the quadrupole by $\Phi$
radians, with minimal dissipation. If the dissipation is rate
dependent, one adds a constraint   that the time of traversal is
$1$.

For viscous media $\gamma=\big(\tau(t),\theta(t)\big)$ is then the
solution of the variational problem
\begin{equation}\label{eq:variation}
{\delta}\int_0^1\left( \frac{ \dot\tau^2 +\dot\theta^2
\sinh^2\tau}2\,-\lambda a(\tau)\sinh(\tau)\dot\theta\right) dt=0
\end{equation}
where $\lambda$ is a Lagrange multiplier and
$\gamma(0)=\gamma(1)=S_0=S(\tau_0,\theta_0)$. This is the
(variation of the) action of a classical particle with charge
$\lambda$ and unit mass moving on the hyperbolic plane (i.e. the
pseudo-sphere) in the presence of a magnetic field $f(\tau)$ given
in Eq.~(\ref{eq:def-f-a}).

Since motion in a magnetic field conserves kinetic energy the
particle moves at constant speed and the dissipation, $\frac 1 2
|\gamma|^2$, depends only on the length of the path. The
variational problem can therefore be cast in purely geometric
terms: Find the shortest closed path starting at a given point
which encloses a given amount of magnetic flux. The shortest path
is evidently also the solution in the case that the dissipation is
rate independent.
%%%%%%%%%

\begin{figure}[h]
\hbox{\includegraphics[width=4cm]{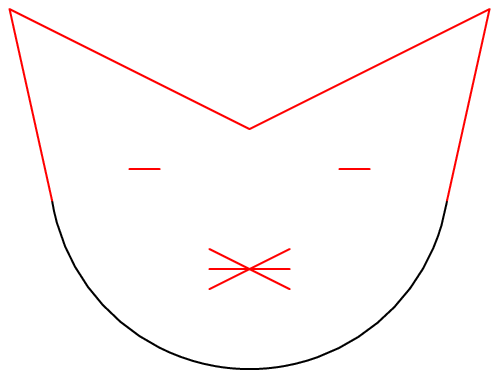}
\includegraphics[width=4cm]{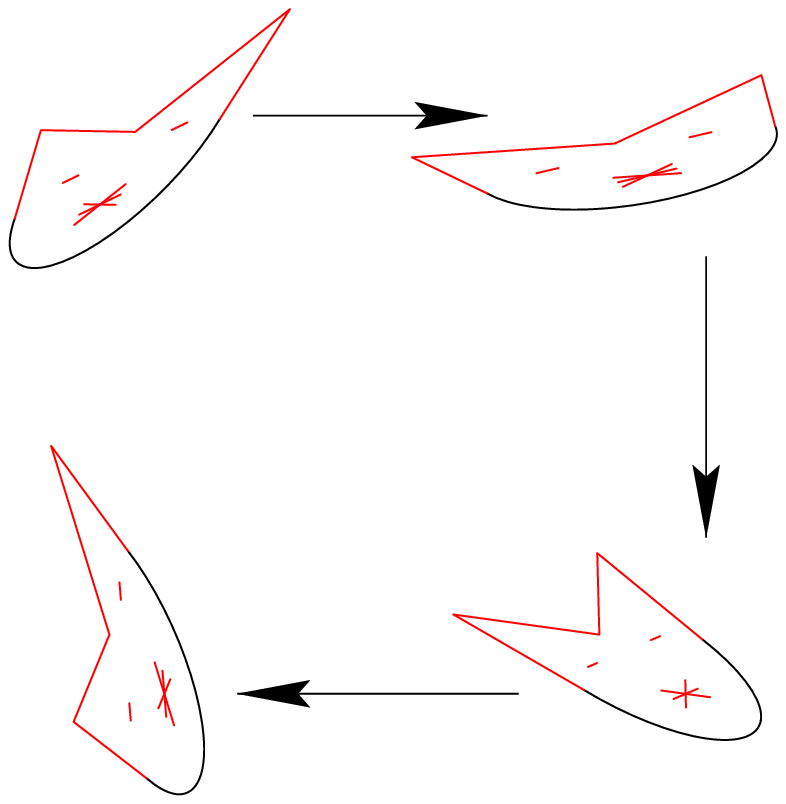}}
\caption{A reference shape (left), assumed to have $Q=1$, and four
instances from the optimal stirring cycle (right).
The configurations on the right are ordered clockwise with
increasing $\theta$, starting with the top left. Both the first and the
last correspond to $\theta=0$ and are therefore related by a pure
rotation.}
\label{fig:chain}
\end{figure}

%%%%%%%%%%%%%%%%%%
Consider the family of isospectral deformations $\tau=const$ which
keep the the eigenvalues of $S$ (or $Q$) constant while rotating
its eigenvectors. We call these ``stirring'', see
Fig.~\ref{fig:chain}.
 One can not stir an isotropic
body, since its eigenvectors don't have well defined directions.

Among the stirring cycles there is an optimal one which maximizes
the rotation per unit length. From
Eqs.~(\ref{eq:CAR2},\ref{eq:metric-2d}) the flux to length ratio
for stirring cycles is $a(\tau)$ of Eq.~(\ref{eq:CAR2}). The
function $a$ takes its maximum at
$\cosh\tau_s=\phi,\tau_s\approx1.061$, see Fig.~\ref{fig:curv}.
Every cycle of the controls rotates by
$\Phi_s=(2-\phi)\pi\approx.382\pi$ radians, somewhat less than a
quarter turn.

\begin{figure}[h]
\includegraphics[width=5cm]{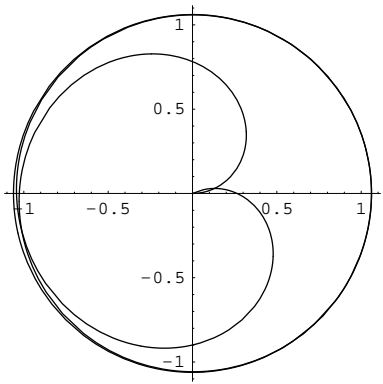}
\caption{The optimal path in the space of shapes
for a rotation by $\pi$ of a body with initial
quadrupole moment $Q=1$. Polar coordinates $(\tau,\theta)$ are
used to parametrize the plane. The
path reaches exponentially close to the optimal stirring cycle at $\cosh\tau=\phi$,
and winds around it twice before returning to the initial configuration.
The optimal stirring cycle is not distinguishable from the envelope of the orbit
in the scale of this figure.}
\label{fig:pi-cat}
\end{figure}

To make full use of the optimal stirring cycle the initial
conditions must be right. This is the case for a quadrupole with
$Q=S(2\tau_s,\theta)$, $\theta$ arbitrary. With other initial
conditions and for large angles of rotations, the optimal paths
approach the optimal stirring cycle, linger near it, eventually
returning to the initial point. This is shown in
Fig.~\ref{fig:pi-cat} for a half turn of an isotropic body.

Since the magnetic field $F(\tau)$ is rotationally invariant, the
angular momentum
\begin{equation}\label{eq:angular-mom}
J=\sinh\tau\left(\dot\theta\sinh\tau - \lambda a(\tau)\right)
\end{equation}
is conserved. Conservation of energy gives
\begin{equation}\label{eq:radial}
0\le\dot\tau^2 =2E- \left(\frac J
{\sinh\tau}+\lambda{a(\tau)}\right)^2,
\end{equation}
which reduces the problem to quadrature. The three conditions,
$\tau(1)=\tau(0)$, $\theta(1)\equiv\theta(0)\,{\rm mod}\ 2\pi$,
and the constraint of enclosed flux $\Phi$, determine the three
parameters $E$, $J$ and $\lambda$.

The initial condition $\tau(0)=0$ is special: By Eq.~(\ref{eq:angular-mom})
the angular momentum $J$ is forced to have the value 0. In turn, there is
also one less condition to satisfy, as the value of $\theta$ when
$\tau=0$ is meaningless. It then follows from Eqs.~(\ref{eq:angular-mom})
and~(\ref{eq:radial}) that the optimal orbit, $\tau(\theta)$, depends
only on the ratio $E/\lambda^2$. Rescaling time properly, we can achieve
$\lambda=1$ by relaxing the constraint that the time to complete
a cycle should be 1.

The key equation controlling the dynamics is Eq.~(\ref{eq:radial}), which
describes effective one-dimensional motion in the potential $a^2(\tau)/2$.
As shown above, $a(\tau)$ has a maximum at $\tau_s$. Therefore, closed orbits
which correspond to optimal paths have energy values $E<a(\tau_s)^2/2$. The
motion is quite simple: Trajectories leave the origin with a positive $\dot\tau$,
reach the turning point $\tau_t=a^{-1}(\sqrt{2E})$, and return symmetrically to
the origin, thereby completing a cycle. The flux accumulated during a cycle
is
\begin{equation}
\Phi=\int_0^{t_{\hbox{\tiny cycle}}}a(\tau)\sinh(\tau)\dot\theta dt\ ,
\end{equation}
where $\dot\theta$ is obtained from Eq.~(\ref{eq:angular-mom}).

Increasingly longer orbits are obtained when $E$ approaches the
separatrix energy $a(\tau_s)^2/2$. The orbit corresponding to the
unstable equilibrium point at $\tau=\tau_s$ is the optimal
stirring cycle, which is a $J=0$ orbit. The flux accumulated
during a complete turn in parameter space, where $\theta$
increases by $2\pi$, is hence bounded by $\Phi_s$. Large values
$\Phi$ require many turns during which the orbit approaches
exponentially the optimal stirring cycle, see
Fig.~\ref{fig:pi-cat}. A circular disc of playdough would
therefore rotate by $\pi$, with no angular momentum, in about
three cycles.

%%%%%%%%%%%%%%%%%%%%%%%%

%%%%%%%%%%%%%%%%%%%%%%%%%%%%%%%%%%

{\bf Acknowledgment:} We thank Amos Ori for useful discussions.
This work is supported by the Technion fund for promotion of
research and by the EU grant HPRN-CT-2002-00277.


\begin{thebibliography}{99}
\bibitem{rem:plane} Note that the plane $t=0$ represents
the single matrix $\sigma_z$ up to rotations and scaling.
\bibitem{ref:ori}We are indebted to Amos Ori for this observation.
\end{thebibliography}
\end{document}